\documentclass[prl,twocolumn,amsfonts,longbibliography]{revtex4-1}
\usepackage{graphicx}
\usepackage{bm}

\begin{document}

\title{Observation of a two-dimensional Fermi gas of atoms}
\author{Kirill Martiyanov}
\author{Vasiliy Makhalov}
\author{Andrey Turlapov}
\email{turlapov@appl.sci-nnov.ru}
\affiliation{Institute of Applied Physics, Russian Academy of Sciences, ul.~Ulyanova 46, Nizhniy Novgorod, 603000, Russia}

\date{\today}

\begin{abstract}
We have prepared a degenerate gas of fermionic atoms which move in two dimensions while the motion in the third dimension is ``frozen'' by tight confinement and low temperature.
{\it In situ} imaging provides direct measurement of the density profile and temperature.
The gas is confined in a defect-free optical potential, and the interactions are widely tunable by means of a Fano--Feshbach resonance.
This system can be a starting point for exploration of 2D Fermi physics and critical phenomena in a pure, controllable environment.
\end{abstract}

\maketitle

Two-dimensional Fermi systems are predicted to have rich physics of phase transitions and quantum critical points~\cite{Sachdev,Korshunov2D2006}.
Reduction of the spatial dimensionality increases the role of fluctuations, which in turn causes phenomena such as superfluidity without Bose-Einstein condensation and non-Fermi-liquid behavior. Reduced dimensionality along with strong interactions is also at the origin of high-temperature superconductivity.
2D collective phenomena have been studied in helium-3 films~\cite{He3Film} and in a rich collection of systems containing electron gas in superconducting~\cite{ScReviewYanase2003,KopaevEng} and nonsuperconducting~\cite{KravchenkoNonSF2DReview,ElectronsOnLiquidHeReview} phases.

An ultracold atomic Fermi gas may provide unique possibilities for studying 2D phenomena in a controlled, impurity-free environment. In 3D atomic Fermi gases~\cite{PitaevskiiFermiReview}, it has been possible to tune at will the principal experimental parameters such as interactions~\cite{Grimmbeta}, energy~\cite{JointScience}, and spin composition~\cite{KetterleImbalanced,HuletPhaseSeparation} and to have more than two spin states~\cite{Jochim3Component2008}. This control and tunability unprecedented to other Fermi systems should also be available in perspective experiments with the 2D ultracold gases.
Experiments with 3D optically trapped atomic Fermi gases~\cite{OHaraScience,AmScientist} and their derivatives have provided first-time observation of fundamental quantum phenomena, which include superfluidity~\cite{Kinast,KinastDampTemp,KetterleVortices,GrimmSFMOI} and mechanical stability~\cite{MechStab} of a resonantly-interacting Fermi gas, coherent transformation of a Fermi system to a Bose system~\cite{Grimmbeta}, Bose-Einstein condensation of molecules~\cite{GrimmBEC,JinBEC,KetterleBEC}, and possibly the shear viscosity near its fundamental quantum minimum~\cite{KazanPaper,ThomasPerfectFluid}.
Three-dimensional ultracold Fermi gases have also been used to test theoretical models of other Fermi systems: neutron stars and nuclear matter~\cite{Baker,Heiselberg}, quark-gluon plasma~\cite{Heinz}, and high-temperature superconductors~\cite{Levin}. Availability of a 2D atomic Fermi gas would allow one to study Fermi mixtures in mixed dimensions, where one species is confined to 2D while the other to 3D~\cite{PetrovMixedD2007,NishidaMixedD2008}. In the field of ultracold gases, Bose systems are experimentally available in three, two, and one spatial dimension~\cite{BlochLowDReview2008}. The Fermi systems, however, have been created only in 3D and 1D~\cite{BlochLowDReview2008,PitaevskiiFermiReview}, while 2D degenerate Fermi systems have been missing.

In this Letter, we report on preparation and {\it in situ} imaging of a 2D atomic Fermi gas.
{\it In situ} observation of a density distribution is among the most direct methods for probing a quantum system. This capability of an atomic gas is unique to presently available 2D Fermi systems. It may let one see phase separation as well as measure thermodynamic, statistical, and mechanical properties, the amount of the mean field, and in some cases, the phase of a many-body wave function.

Two-dimensional kinematics is obtained by confining atoms in a highly anisotropic pancake-shaped potential
\begin{equation}
V(\vec{x})=\frac{m\omega_z^2z^2}2+\frac{m\omega_\perp^2(x^2+y^2)}2,\quad\omega_z\gg\omega_\perp
\label{eq:Potential}
\end{equation}
and keeping their energy below the energy of the first axial excited state.

\begin{figure}[htb!]
\begin{center}
\includegraphics[width=2.7in]{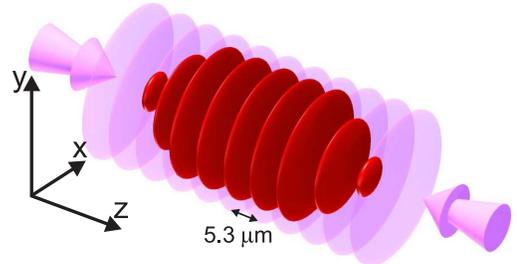}
\end{center}
\caption{Trapping ultracold atoms in antinodes of a standing optical wave. The isolated clouds of atoms shown in dark red, the standing-wave intensity shown in light purple.}
\label{fig:ExpSetup}
\end{figure}

\begin{figure}[htb!]
\begin{center}
\includegraphics[width=3.3in]{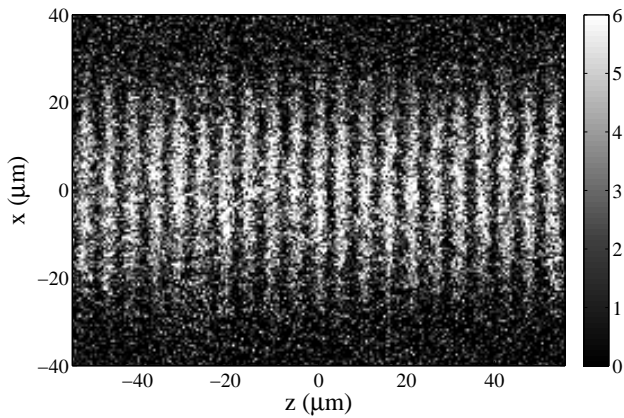}
\end{center}
\caption{{\it In situ} image of the column density distribution $n_2(x,z)$ of the Fermi gas trapped in antinodes of a standing optical wave. Each cloud is a 2D system. The number of atoms per spin state per square micron is coded in the tones of gray. The meaning of the tones is shown on the right.}
\label{fig:Shadow}
\end{figure}

A series of pancake-shaped potentials is created by setting up a standing optical wave along the $z$ direction as shown in Fig.~\ref{fig:ExpSetup}.
The frequency of the standing wave is far below the resonance of the atoms. As a result, the minima of the dipole potential are at the intensity maxima and each antinode acts on the atoms nearly as the potential of Eq.~(\ref{eq:Potential}). The Gaussian shape of the mode assures weak confinement along the transverse directions. In Fig.~\ref{fig:Shadow}, one may see a snap shot of the density distribution taken along the $y$ direction, parallel to the plane of the pancake-shaped clouds.
The gas is prepared in a strongly degenerate regime, with the temperature of $\simeq 0.1\,E_F$, where $E_F$ is the Fermi energy. The interactions are tunable by means of the Fano--Feshbach resonance. This system can be used for qualitative and quantitative tests of many-body theories in two dimensions.

In experiment, we use the atoms of lithium-6 in the two lowest-energy spin states, $|1\rangle$ and $|2\rangle$, with about equal populations: $50\pm 2.5\%$. In the limit of high magnetic field, these states have the same electronic spin projection: $m_j=-1/2$ in the magnetic-field basis, and different projections of the nuclear spin, $m_I=1$ and $m_I=0$, respectively. When modeling condensed matter systems in experiments with cold atoms, states $|1\rangle$ and $|2\rangle$ may be regarded as analogs of the electronic spin-up and spin-down states, respectively.

The standing-wave dipole trap is formed by two focused, counterpropagating Gaussian beams with overlapping foci. The beams have identical power and polarization, and the wavelength of 10.6 $\mu$m. The period of the dipole potential is 5.3 $\mu$m, which can be resolved on images taken in the light at 671 nm wavelength, which is resonant to the lithium atoms.

The degenerate Fermi gas is prepared following the procedure of Ref.~\cite{FermionsInLattice}.
During the first 6 s of preparation, $10^8-10^9$ atoms are collected in a magneto-optical trap (MOT) from an atomic beam. The standing-wave optical dipole trap is spatially overlapping with the MOT and is on during the MOT loading. After the MOT fields are turned off, $\sim 10^6$ atoms remain trapped in the standing-wave dipole trap whose depth is $\simeq 230$ $\mu$K. In order to start evaporative cooling, the $s$-wave scattering length is increased to a large negative value $a=-3950$ Bohr by switching on a nearly uniform magnetic field of 1020 G in the $-y$ direction. This value of the magnetic field corresponds to the Fermi side of a broad Fano--Feshbach resonance in $s$-wave scattering~\cite{BartensteinFeshbach}. During 1 s, the gas evaporates freely in a stationary potential. Afterwards, one of the beams forming the standing wave is gradually turned off over 0.2 s. As a result, the gas adiabatically reloads to a cigar-shape dipole trap formed by a focus of a single traveling wave. During reloading the gas keeps evaporating and cooling. Further cooling is done by means of forced evaporation~\cite{LeCooling} during $10.6$ s. Over this time, the trap depth is decreased by a factor of 100 following the law $U_1(t)=(60\mbox{ $\mu$K})\times\left[1-t/(14\mbox{ s})\right]^{3.24}$ by decreasing the trapping-beam power.
In the single-beam dipole trap at low depth, the axial confinement is dominated by the small curvature of the magnetic field, which compresses the cloud in the axial direction. At this point, the second beam is reestablished during 0.2 s, which reloads the gas back into the standing-wave trap.
Forced evaporation is then continued by decreasing the depth exponentially for 3.5 s by a factor of 10. Afterwards, the trap depth is kept stationary for 0.5 s. Then the power of the trapping beams adiabatically, over 2.4 s, increases to get a potential of desired height. This completes the preparation.

{\it In situ} images of the gas are obtained by the absorption imaging technique~\cite{Kinast}: The atoms are irradiated by a 6 $\mu$s pulse of a uniform laser beam resonant to a cycling two-level transition for one of the two spin states. The imaging beam is shed in the $y$ direction, which is opposite to the magnetic field and perpendicular to the axis of the trap cylindrical symmetry $z$. The shadow, which the atoms make in the imaging beam, is projected and recorded on a CCD camera. From the shadow, we reconstruct the column density distribution $n_2(x,z)$ exactly accounting for saturation effects~\cite{FermionsInLattice} due to finite intensity of the imaging beam: $I=2.6\mbox{ mW/cm}^2\simeq I_{\mbox{sat}}$.
In counting the atom number, we correct for the effects that reduce the visible number of atoms:
(i) Doppler shift caused by the acceleration due to the light-pressure force, (ii) diffractive spreading of the image of narrow clouds slightly beyond the objective size, (iii) small coupling of the imaging transition to a dark state, and (iv) fluorescence. These effects reduce the apparent atom number by 18\%, 8\%, 3\%, and 3\% respectively.
The spatial resolution of our video system is $1.2-2.0$ $\mu$m. The observed density distribution is shown in Fig.~\ref{fig:Shadow}. The noise is dominated by the photon shot noise of the imaging beam. Each cloud in the figure represents an isolated two-dimensional Fermi system.

The confining potential is a combination of a deep optical lattice in the $z$ direction and Gaussian-shape confinement in the transverse plane:
\begin{equation}
V_s(\vec{x})=sE_r \left(1-\exp\left[-\frac{m\omega_\perp^2(x^2+y^2)}{2sE_r}\right]\,\cos^2kz\right),
\label{eq:OptLattice}
\end{equation}
where $E_r=\hbar^2k^2/2m$ is the recoil energy [$k=2\pi/(10.6\mbox{ $\mu$m})$] and $s$ is the dimensionless lattice depth. In formula~(\ref{eq:OptLattice}) for the potential shape, we neglect beam divergence  because the Rayleigh length $z_R\simeq 5$ mm is much bigger than the 400 $\mu$m long region where the atoms are prepared. At the bottom of each well, the potential is nearly harmonic as in Eq.~(\ref{eq:Potential}). We use a version of the parametric resonance method~\cite{HaenschParametric} to measure the frequencies $\omega_\perp/2\pi=102\pm 4$ Hz and $\omega_z/2\pi=5570\pm 100$ Hz and the dimensionless depth $s=(\hbar\omega_z/2E_r)^2=86.5\pm 3.0$. The absolute trap depth value is $sE_r=(4.65\pm 0.18)\hbar\omega_z=1.23$ $\mu$K. Each trap is strongly anisotropic: $\omega_z/\omega_\perp=54.6\pm 2.4$.

Reduced dimensionality of the gas in a single cell can be proven by satisfying both of the following conditions: (i) The absolute majority of atoms is populating the axial ground state and (ii) tunneling between the wells is negligible.

The suppression of tunneling is assured by the lattice depth $s=86.5\gg 1$. The width of the lowest Bloch band is $1.1\times 10^{-7}\hbar\omega_z$, which gives the tunneling time of $\simeq 260$ s. For times much shorter than this, the gas remains kinematically two-dimensional.

The population of the excited axial states may come from the Fermi statistics, thermal excitations, and mean-field interaction. In estimating the role of the statistics and temperature, we use the model of noninteracting Fermi gas in a parabolic potential.

At zero temperature, due to the Pauli exclusion principle, atoms occupy energy levels up to the Fermi energy $E_F=\hbar\omega_\perp\sqrt{2N}$, where $N$ is the number of atoms per spin state in each cell and $E_F$ is counted from the axial ground state. The necessary condition of dimensionality two is then $E_F<\hbar\omega_z$. We find $N=660\pm 60$ by integrating the column density in each cell, where the error margins include deviations in the measurement between different repetitions of the experiment as well as fluctuations between different clouds on the same photograph (Fig.~\ref{fig:Shadow}). For the analysis we use 21 central clouds which are about equally populated.
Therefore,
$E_F=\hbar\omega_\perp\sqrt{2N}=180\pm 10\mbox{ nK}=(0.67\pm 0.04)\,\hbar\omega_z$: i.~e., the Pauli exclusion principle does not create population of the axial excited state.

\begin{figure}[htb!]
\begin{center}
\includegraphics[width=3.0in]{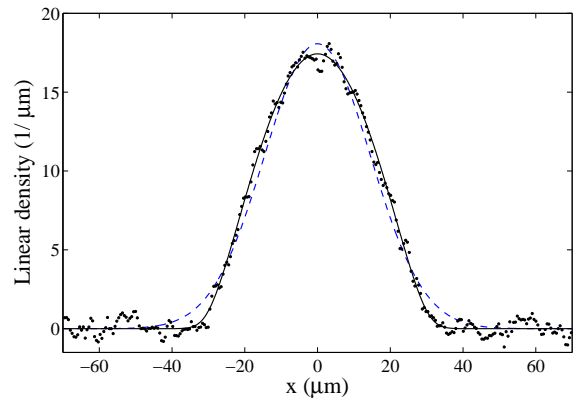}
\end{center}
\caption{One-dimensional density profile in a single cloud obtained by integrating 3D density distribution along $y$ and $z$. The dots are the data averaged over the 21 central clouds. The black solid curve is the fit of formula~(\ref{eq:Profile1D}) to the data. The blue dashed curve is the fit of a Gaussian distribution.}
\label{fig:Transverse}
\end{figure}

The temperature is found by fitting the model one-dimensional density profile to the data. Figure~\ref{fig:Transverse} shows the experimental density distribution obtained by integrating $n_2(x,z)$ along $z$ in a single cell and averaging over 21 central cells. Also for the data of Fig.~\ref{fig:Transverse}, traveling average over adjacent pixels is done to reduce the photon shot noise effect.
The model density profile is the finite-temperature Thomas--Fermi distribution
\begin{equation}
n_1(x)=-\sqrt\frac{m\omega_\perp}{2\pi\hbar}\left(\frac{T}{\hbar\omega_\perp}\right)^{3/2} \mbox{Li}_{3/2}\left(-e^{\frac{\mu}{T}-\frac{m\omega_\perp^2x^2}{2T}}\right),
\label{eq:Profile1D}
\end{equation}
where $\mbox{Li}_{3/2}$ is the polylogarithm function of order 3/2 and $\mu$ is the chemical potential found from the condition $N=\int dx\,n_1(x)$. The fitting parameters are $T$ and $\omega_\perp$. We make $\omega_\perp$ floating in the fit despite the availability of a measured value in order to account for the presence of a small mean field. The fitted curve is shown in Fig.~\ref{fig:Transverse}.
Fitting of a Gaussian (blue dashed curve), which would correspond to a nondegenerate gas, gives significantly larger deviation from the data.
Fitting of the Thomas--Fermi profile~(\ref{eq:Profile1D}) yields $T=(0.10\pm 0.03)E_F=18$ nK. At this temperature, just 0.01\% of the atoms are thermally excited out of the axial ground state.

The mean field can be treated as a perturbation because the 3D interaction parameter $k_Fa=-0.43$ is small ($k_F=\sqrt{2mE_F}/\hbar$) and because the system is far from the predicted geometric resonance~\cite{Shlyapnikov2DResonance}. We estimate the depletion of the axial ground state by using a simplified Hamiltonian for relative motion of two colliding particles in the center-of-mass reference frame:
\begin{equation}
\hat{H}=\frac{\hat{P}_z^2}{2\bar{m}}+\frac{\bar{m}\omega_z^2\hat{Z}^2}{2}+\frac{2\pi\hbar^2a\,\langle n_2(x,y)\rangle}{\bar{m}}\delta(\hat{Z}).
\end{equation}
where $\bar{m}=m/2$ is the reduced mass; $\hat{P}_z$ and $\hat{Z}$ are the operators of the relative momentum and coordinate of the two interacting atoms, respectively; $\delta$ is the Dirac delta function; and $\langle n_2(x,y)\rangle=mE_F/3\pi\hbar^2$ is the trap-averaged 2D density distribution in the transverse plane. In this approximation, only the motion along $z$ is accounted for. The last term accounts for the contact interaction, which depletes 0.2\% of the atoms out of the axial ground state.

As a result, the total depletion of the axial ground state is estimated to be $\simeq 0.2\%$. Together with the negligible tunnel rate, this proves that the axial motion is ``frozen out,'' and we observe a series of isolated 2D Fermi systems.

The technique of trapping fermionic atoms in antinodes of a standing wave has been used in the beautiful experiments devoted to molecular formation~\cite{Grimmmolec}, interferometry~\cite{InguscioFermiInterferometer}, and study of collisions~\cite{Esslinger2D,Thomas2D}. None of these experimental systems have been found to be in the 2D Fermi-degenerate regime. In particular, in experiment~\cite{Esslinger2D}, the gas is not two-dimensional because the preparation process inevitably populates at least the two lowest Bloch bands and also because the tunnel time is shorter than the interval between preparation and probing.
In experiment~\cite{Thomas2D}, the dimensionality of the lowest-energy samples is not clear and represents an interesting theoretical question.
On one hand, the Fermi energy is above the first axial excited state. On the other hand, due to the strong interactions, the zero-temperature chemical potential could be still below $\hbar\omega_z$. The thermal populations of the excited axial states are unknown but can, in principle, be determined from the reported data via an appropriate theoretical analysis. To our knowledge, such analysis is not available at present. If in experiment~\cite{Thomas2D} the thermal populations are negligible, that system might be in the interesting quasi-two-dimensional regime, where the axial excited states are populated by the effect of the strong interactions alone~\cite{Duan2DDepletion}.

In conclusion, we have prepared and directly observed a two-dimensional Fermi gas of atoms. This system may be a starting point for exploration of 2D Fermi physics and critical phenomena in a defect-free, controllable environment.

\begin{acknowledgments}
We are thankful to Yu.~M.~Kagan and V.~S.~Babichenko of RRC Kurchatov Institute for stimulating discussions. We thank A.~G.~Litvak for continuous interest and support.
We acknowledge the financial support by the National Project ``Education,'' programs of the Presidium of Russian Academy of Sciences ``Quantum physics of condensed media'' and ``Nonlinear dynamics,'' and Russian Foundation for Basic Research (Grants No. 08-02-01249-a and No. 08-02-01821-e\_b). A.T. acknowledges a stipend from the Russian Science Support Foundation.

\end{acknowledgments}

\end{document}